\documentstyle[aps,pre]{revtex}

\begin{document}

\draft

\title{Approach to Quantum Kramers' Equation and Barrier Crossing Dynamics}

\author{Dhruba Banerjee$^1$, Bidhan Chandra Bag$^2$, Suman Kumar Banik$^1$
and Deb Shankar Ray$^1${\footnote{e-mail: pcdsr@mahendra.iacs.res.in}}
}

\address{
$^1$Indian Association for the Cultivation of Science,
Jadavpur, Kolkata 700 032, India.\\
$^2$Department of Chemistry, Visva-Bharati, Shantiniketan 731 325, India.
}

\date{\today}

\maketitle

\begin{abstract}
We have presented a simple approach to quantum theory of Brownian motion
and barrier crossing dynamics.
Based on an initial coherent state representation of bath oscillators and an 
equilibrium canonical distribution of quantum mechanical mean values of their
co-ordinates and momenta we have derived a $c$-number generalized
quantum Langevin equation. The approach allows us to implement the method of classical
non-Markovian Brownian motion to realize an exact generalized 
non-Markovian quantum Kramers' equation.  
The equation is valid for arbitrary temperature and friction. We have solved
this equation in the spatial diffusion-limited regime to derive quantum
Kramers' rate of barrier crossing and analyze its variation as a function
of temperature and friction. 
While almost all the earlier theories rest on quasi-probability distribution 
functions (like Wigner function) and path integral methods, 
the present work is based on 
{\it true probability distribution functions} 
and is independent of path integral
techniques. The theory is a natural extension of the classical theory to
quantum domain and provides a unified description of thermal 
activated processes and tunneling.
\end{abstract}

\pacs{PACS number(s): 05.40.-a, 02.50.-r}

\section{Introduction} 
Ever since Kramers \cite{kram} reported his seminal work on the Brownian motion 
in phase space, the theory of noise-induced escape from metastable states 
has become a central issue in several areas of physical and chemical
sciences. The
escape is governed by Brownian motion in addition to the characteristic 
dynamical motion of the system in presence of a potential $V(X)$, Brownian
motion being due to the thermal forces, which in turn are associated with the
dissipation through fluctuation-dissipation relation at a finite temperature 
$T$. The problem and many of its variants have been addressed by a large number 
of workers over the last several decades at various levels of description
and have been extended to semi-classical and quantum 
domains \cite{kram,hangg,mel,tani,tani1,stru,joac,laio,mant,bier,ray,ray1,ray2,ray3}. 
A major impetus in the development of quantum theory of 
dissipative processes was the discovery of laser in sixties followed by
significant advancement in the field of nonlinear and quantum optics in 
seventies and eighties when the extensive applications of nonequilibrium 
quantum statistical methods were made \cite{loui,gang}. Various nonlinear optical processes and
phenomena were described with the help of operator Langevin equations, density 
operator methods and the associated quasiclassical distribution functions.
However these dynamical semigroup methods of quantum optics could not gain much 
ground in the
theory of activated rate processes due to the fact that these are primarily 
based on system-reservoir weak coupling and Markov approximations 
\cite{loui}, which are often
too drastic in the situations pertaining to chemical dynamics 
and condensed matter physics.
Subsequent to these developments quantum Brownian motion 
\cite{hangg,tani,tani1,mill,cao,liao,legg,grab,legg1} emerged as a subject
of renewed interest in early eighties when the problem of dissipative quantum
tunneling was addressed by Leggett and others and almost simultaneously
quantum Kramers' problem and some allied issues 
attracted serious attention of a number of physical
chemists. We refer to \cite{hangg,mel,talk} for an overview.

The aforesaid development of the quantum theory of Brownian motion essentially 
rests on the method of functional integrals \cite{legg}. This is based on the 
calculation of the partition function for the Hamiltonian of the system  
coupled to its environment and a non-canonical quantization procedure. The 
classical theories on the other hand rely on the partial differential 
equations describing the evolution of the probability distribution functions
of the system both in the Markovian and non-Markovian regions
\cite{kram,mel,adel,mazo,hang,grot}. The methods of
the classical and the quantum theories are thus widely different in 
their approaches. The question is whether there is any natural extension of the 
classical
theory to the quantum domain. For example, one might ask what is the quantum 
analogue of the classical Kramers'  equation or Smoluchowskii's equation ? Is 
it possible to generalize Kramers' method of treatment of barrier 
crossing dynamics within a quantum mechanical framework so that the usual 
classical thermal activated process and
quantum tunneling can be described within an unified scheme?

We intend to address these issues in the present paper. Specifically our
object is two fold:

1) Our primary aim here is to develop a {\it quantum analogue of classical
non-Markovian Kramers' equation}, which describes quantum Brownian motion
of a particle in a force field at arbitrary temperature and coupling
in terms of {\it true probability distribution} function rather than 
quasi-probability function \cite{epw,qpdf}. The
generalized quantum Kramers' equation (GQKE) reduces to its classical 
counterpart in the limit $\hbar \rightarrow 0$ both in Markovian and 
non-Markovian descriptions. The probability distribution functions remain
well-behaved in the full quantum limit.

2) While the existing methods of calculation of quantum Kramers' rate are 
based on path integral techniques, we solve GQKE 
for barrier crossing dynamics as a boundary value problem taking care
of the full quantum nature of the system and the bath. The generalized rate 
in the spatial-diffusion-limited regime
reduces to  Kramers-Grote-Hynes' rate in the classical limit and to pure
tunneling rate in the quantum limit at zero temperature. To the best of our 
knowledge the implementation of a differential equation based approach like 
the present one  
has not been tried till date for a full quantum mechanical calculation of the
rate.

Based on an initial coherent state representation of the bath oscillators  and 
an equilibrium canonical distribution of the quantum mechanical mean values
of their co-ordinates and momenta, we derive a generalized quantum Langevin 
equation in $c$-numbers.
The bath imparts classical looking c-number quantum
noise which satisfies standard fluctuation-dissipation relation. The simplicity
of the treatment lies in the fact that the quantum Langevin equation is 
amenable to a theoretical analysis in terms of the well known classical
non-Markovian theory of Brownian motion \cite{adel,mazo} so that the GQKE assumes the  
form of its classical counterpart.
In what follows we show that
GQKE can be solved in the spirit of Kramers' method to calculate the quantum 
escape rate. This recasting of the quantum problem into a classical form
allows us to realize the various limits of the theory on a general footing. 

The organization of the paper is as follows. We introduce the system-reservoir
model and an ensemble averaging procedure to derive the generalized 
quantum Langevin equation in a c-number form in the next section.
In section III we derive the corresponding GQKE followed by a calculation
of quantum rate of escape in section IV. The key result is illustrated in 
section V by assuming a Lorentzian density distribution of bath
oscillators and a specific cubic
potential. The paper is concluded in section VI.

\section{The generalized Quantum Langevin equation in $c$-numbers}

To start with we consider the standard system-heat bath model of Zwanzig
form \cite{zwanzig}. The Hamiltonian is given by
\begin{equation}
\label{eq1}
\hat{H} = \frac{\hat{P}^2}{2} + V(\hat{X}) + \sum_j \left [
\frac{\hat{p}_j^2}{2} + \frac{1}{2} \kappa_j ( \hat{q}_j - \hat{X} )^2
\right ] \; \; .
\end{equation}

\noindent
Here $\hat{X}$ and $\hat{P}$ are the co-ordinate and momentum operators
of the Brownian particle (the system) of unit mass and the set
\{$\hat{q}_j, \hat{p}_j$\} is the set of co-ordinate and momentum operators
for the heat bath particles. The mass of the $j$-th particle is unity
and $\kappa_j$ is the spring constant of the spring connecting it
to the Brownian particle. The potential $V(\hat{X})$ is due to the external 
force
field for the Brownian particle. The co-ordinate and momentum operators
follow the usual commutation relation
\begin{equation}
\label{eq2}
[\hat{X},\hat{P}] = i\hbar \; \; \; {\rm and} \; \; \;
[ \hat{q}_j, \hat{p}_j ]  = i \hbar \delta_{ij} \; \; .
\end{equation}

Eliminating the bath degrees of freedom in the usual way
\cite{loui,bjw} we obtain the 
operator Langevin equation for the particle,
\begin{equation}
\label{eq3}
\ddot{\hat{X}} + \int_0^t dt' \beta (t - t') \dot{\hat{X}} (t')
+ V'(\hat{X}) = \hat{F} (t) \; \; ,
\end{equation}

\noindent
where the noise operator $\hat{F}(t)$ and memory kernel $\beta(t)$ are given 
by
\begin{equation}
\label{eq4}
\hat{F}(t) = \sum_j [
\{ \hat{q}_j(0) - \hat{x} (0) \}
{\kappa}_j \cos \omega_j t + \hat{p}_j(0) 
{\kappa}_j^{\frac{1}{2}} 
\sin \omega_j t ]
\end{equation}

\noindent
and

\begin{equation}
\label{eq5}
\beta (t) = \sum_j {\kappa}_j \cos \omega_j t
\end{equation}

\noindent
where $\omega_j^2 = \kappa_j$ and the initial variations of the 
heat bath variables $\hat{q}_j(0)$ and $\hat{p}_j(0)$ 
occur in the force term $\hat{F}(t)$.
The relevant quantum statistical average are well known

\begin{equation}
\label{eq6}
{\langle \hat{F} \rangle}_{QS} = 0
\end{equation}

\noindent
and

\begin{equation}
\label{eq7}
\frac{1}{2} \left[{\langle \hat{F}(t') \hat{F}(t)\rangle}_{QS} + 
{\langle \hat{F}(t) \hat{F}(t')\rangle}_{QS}\right] = \frac{1}{2} 
\sum_j {\kappa}_j \hbar \omega_j (\coth \frac{\hbar \omega_j}{2k_bT}) \cos \omega_j
(t-t') 
\end{equation}

\noindent
Here $\langle \cdots \rangle_{QS}$ refers to quantum statistical average on the bath
degrees of freedom. To arrive at the above relations one assumes \cite{bjw}
that the bath
oscillators are canonically distributed with respect to the bath 
Hamiltonian at $t=0$ so that for any operator $\hat{O}$ the average is 

\begin{equation}
\label{eq8}
{\langle \hat{O} \rangle}_{QS} = 
\frac{ {\rm Tr}\; \hat{O} \; \exp \left ( -\hat{H}_{bath} / k_bT \right ) 
}{ 
{\rm Tr} \; \exp \left ( - \hat{H}_{bath} / k_bT  \right ) }
\end{equation}

\noindent
where $\hat{H}_{bath} = \sum_j [ ( \hat{p}_j^2 / 2 ) + (1/2) \kappa_j
\{ \hat{q}_j - \hat{x} \}^2 ]$
By trace we mean carrying out quantum statistical average with number states of the
bath oscillators multiplied by the arbitrary state of the particle. 
Eq.(\ref{eq7}) is the celebrated fluctuation-dissipation relation (FDR).

Eq.(\ref{eq3}) is an exact quantum Langevin equation in operator form which is a 
standard textbook material \cite{loui,bjw}. Our aim here is to replace it by an 
equivalent quantum generalized Langevin equation (QGLE) in $c$-numbers. 
It is important to mention here that again
this is not a new problem \cite{loui,gang} so long as one is restricted to 
standard quasi-classical methods of Wigner functions and the like.
In general, however, one is confronted with serious trouble of 
negativity or singularity of these quasi-probability distribution
functions in the full quantum domain. To address the problem of quantum
non-Markovian dynamics in terms of a {\it true probabilistic description}
we, however, follow a different procedure.
Our approach here is to split up the quantum statistical averaging procedure in
{\it two distinct} steps. We {\it first} carry out the quantum mechanical
average of Eq.(\ref{eq3})

\begin{equation}
\label{eq9}
\langle \ddot{\hat{X}} \rangle +\langle V'(\hat{X})\rangle 
+ \int_0^t \beta (t-t') \langle \dot{\hat{X}}(t') \rangle dt'
= \langle \hat{F}(t)\rangle
\end{equation}

\noindent
where the averaging is taken over the initial product separable quantum states
of the particle and the bath oscillators at $t=0$, 
$|\phi \rangle \{|\alpha_1\rangle |\alpha_2\rangle  \cdots |\alpha_N\rangle \}$.
Here $|\phi \rangle$ denotes any arbitrary initial state of the particle and
$|\alpha_i\rangle$ corresponds to the initial coherent state of the i-th
bath oscillator. $|\alpha_i\rangle$ is given by
$| \alpha_i \rangle = \exp (- | \alpha_i |^2 / 2 )
\sum_{n_i=0}^{\infty} ( \alpha_i^{n_i} / \sqrt{ n_i } ) | n_i \rangle$,
$\alpha_i$ being expressed in terms
of the variables of the coordinate and momentum of the $i-$th oscillator
$\langle \hat{q}_i (0)\rangle = 
\sqrt{ \hbar / (2\omega_i) } \left(\alpha_i+ \alpha_i^*\right)$ 
and $\langle \hat{p}_j (0) \rangle = i 
\sqrt{ \hbar \omega_i / 2} (\alpha_i - \alpha_i^*)$,
respectively. It is important to note that $\langle \hat{F}(t) \rangle$ of
Eq.(\ref{eq9}) is a classical-like noise term which, in general, is a non zero number
because of the averaging procedure over the coordinate and momentum operators
of the bath oscillators with respect to initial coherent state and is given by

\begin{equation}
\label{eq10}
\langle \hat{F}(t) \rangle = 
\sum_j \left [ 
\{ \langle \hat{q}_j (0)\rangle - \langle \hat{x} (0)\rangle \}
\kappa_j 
\cos w_j t +
\langle \hat{p}_j(0)\rangle \kappa_j^{1/2} \sin w_j t \right] = f(t), 
\rm{say} \; \; .
\end{equation}
 
\noindent
We now turn to the {\it second} averaging. 
To realize $f(t)$ an effective $c$-number noise we now demand that it 
must satisfy 

\begin{eqnarray*}
{\langle f(t) \rangle}_s =0
\end{eqnarray*}

and

\begin{equation}
\label{eq11}
{\langle f(t) f(t') \rangle}_s = \frac{1}{2} 
\sum \kappa_j \hbar \omega_j (\coth \frac{\hbar \omega_j}{2k_bT}) \cos\omega_j
(t-t') 
\end{equation}

That is, $f(t)$ is zero centered and satisfies quantum fluctuation-dissipation
relation. This may be achieved if and only if one introduces the following 
canonical distribution of quantum mechanical mean values of the bath oscillators
at $t=0$,

\begin{equation}
\label{eq12}
P_j = \exp \left[ - 
\frac{ [ \omega_j^2 
\{ \langle \hat{q}_j (0) \rangle - \langle \hat{x} (0) \rangle \}^2 
+ {\langle \hat{p}_j (0) \rangle}^2]}{2 \hbar \omega_j (\bar{n}_j 
+ \frac{1}{2})}\right]
\end{equation}

\noindent
so that for any quantum mechanical mean value 
$O_j (\langle \hat{p}_j(0) \rangle, 
\{ \langle \hat{q}_j(0) \rangle - \langle \hat{x} (0) \rangle \} )$,
the statistical average is

\begin{equation}
\label{eq13}
\langle O_j \rangle_S = \int O_j(\langle \hat{p}_j(0) \rangle, 
\{ \langle \hat{q}_j(0) \rangle - \langle \hat{x} (0) \rangle \} ) \; 
P_j \; d\langle \hat{p}_j(0) \rangle \;
d \{ \langle \hat{q}_j(0) \rangle - \langle \hat{x} (0) \rangle \}
\end{equation}

Here $\bar{n}_j$ indicates the average thermal photon number of the $j$-th oscillator
at temperature $T$ as defined by
$\bar{n}_j = 1/[ \exp ( \hbar \omega_j / k_BT ) - 1 ]$.

To proceed further we now add the force term $V'(\langle \hat X \rangle)$ on 
both sides of Eq.(\ref{eq9}) and rearrange it to obtain formally
\begin{equation}
\label{eq14}
\langle \ddot{\hat{X}} \rangle + V'(\langle \hat{X}\rangle)+ \int_0^t \beta (t-t')  
\langle \dot{\hat{X}} (t') \rangle dt'
=f(t) + Q(t)
\end{equation}

\noindent
where 
\begin{equation}
\label{eq15}
Q(t) = V'(\langle \hat{X} \rangle) - \langle V'(\hat{X})\rangle
\end{equation}

\noindent
represents the quantum mechanical dispersion of the force operator $V'(\hat{X})$
due to the system degree of freedom. Since $Q(t)$ is a quantum
fluctuation term Eq.(\ref{eq14}) offers a simple interpretation. This implies that the
classical looking generalized quantum Langevin equation is governed by a $c$
-number quantum
noise $f(t)$ which originates from the quantum mechanical heat bath characterized
by the properties (\ref{eq11}) and a quantum fluctuation term $Q(t)$ due to the 
quantum nature of the system characteristic of the nonlinearity of the potential.
$Q(t)$ can be calculated order by order. In Appendix-A we show how $Q(t)$
can be calculated in the lowest order.

Summarizing the above discussions we point out that it is possible to formulate
a QGLE (\ref{eq14}) of the quantum mechanical mean value of the coordinate of the
Brownian particle in a field of potential $V(\hat{X})$, provided a
classical-like noise term $f(t)$ due to thermal bath satisfies (\ref{eq11}) where the
ensemble average has to be carried out with distribution (\ref{eq12}). To realize $f(t)$
as a noise term we have split up the quantum statistical averaging ${\langle \cdots \rangle}_{QS}$
into a quantum mechanical mean ${\langle \cdots \rangle}$ by the explicit use of an initial coherent
state representation of the bath oscillators and then a statistical averaging
${\langle \cdots \rangle}_s$ of the quantum mechanical mean values with (\ref{eq12}).
 It is easy
to note that the distribution of the quantum mechanical mean values of the bath
oscillators (\ref{eq12}) reduces to classical Maxwell-Boltzmann distribution in the thermal
limit $\hbar \omega_j \ll k_bT$, i.e.
$\exp [ - ( \omega_j^2 
\{ \langle \hat{q}_j (0) \rangle - \langle \hat{x} (0) \rangle \}^2
+ \langle \hat{p}_j (0) \rangle^2 ) / 2 k_BT ]$.
Secondly, 
the vacuum term in the distribution (\ref{eq12}) prevents the distribution
function from
being singular at $T=0$. In other words the width of distribution remains
finite even at absolute zero, which is a simple consequence of uncertainty
principle.

\section{The generalized quantum Kramers' equation}

It is now convenient to rewrite the generalized Langevin equation (\ref{eq14}) of the
Brownian particle in presence of an external force field in the form

\begin{equation}
\label{eq16}
\ddot{x}  + V'(x) + \int_0^t \beta (t-t')  \dot{x}(t') dt'=f(t) + Q(t)
\end{equation}

\noindent
where we let $\langle \hat{X} \rangle = x$ for a simple notational change. $\beta(t)$
is the dissipative kernel and $f(t)$ is the zero-centered stationary noise due 
to the reservoir where

\begin{equation}
\label{eq17}
{\langle f(t) \rangle}_s  =0 , \; \; \;
{\langle f(t) f(t')\rangle}_s =c(|t-t'|) = c(\tau)
\end{equation}

\noindent
Here $c(\tau)$ is the correlation function which in the equilibrium state is
connected to memory kernel $\beta(t)$ through FDR of the form

\begin{equation}
\label{eq18}
c(t-t') = \frac{1}{2} \int_0^\infty d\omega \kappa(\omega) \rho(\omega) \hbar 
\omega\left[\coth \frac{\hbar \omega}{2 k_b T}\right] \cos {\omega}(t-t')
\end{equation}

\noindent
Eq.(\ref{eq18}) is the continuum version of Eq.(\ref{eq11}). $\rho(\omega)$ is the density
of modes of the reservoir oscillators. In the continuum version $\beta(t)$ is given
by

\begin{equation}
\label{eq19}
\beta (t-t') = \int_0^\infty d\omega \kappa(\omega) \rho(\omega) \cos \omega
(t-t')
\end{equation}

\noindent
In the high temperature limit $(k_bT>> \hbar \omega)$ one recovers the well-known
classical FDR through

\begin{equation}
\label{eq20}
c(t-t') = k_bT \beta (t-t')
\end{equation}

We now proceed to the solution of Eq.(\ref{eq16}). One of the essential step in this direction
is to linearize the potential $V(x)$ in the left hand side of Eq.(\ref{eq16}) around 
the bottom of the well at
$x= x_0$ so that $V(x) = V(x_0) + (1/2) \omega_0^2 (x-x_0)^2$. 
$\omega_0^2$ refers to the second derivative of the potential $V(x)$ evaluated
at $x=x_0$. This together
with a Laplace transform of Eq.(\ref{eq16}) leads us to the following general 
solution (we take $x_0=0$ for the present section)

\begin{equation}
\label{eq21}
x(t) = \langle x(t) \rangle_s +\int_0^t M_0(t-\tau) f(\tau) d\tau
\end{equation}

\noindent
where

\begin{equation}
\label{eq22}
\langle x(t)\rangle_s = v(0) M_0(t) + x(0) \chi_x(t) + G_0(t)
\end{equation}

\noindent
and

\begin{equation}
\label{eq23}
G_0(t) =\int_0^t M_0(t-\tau) Q_0(\tau) d\tau \; \; ,
\end{equation}

\begin{equation}
\label{eq24}
\chi_x(t) = 1- \omega_0^2 \int_0^t M_0(\tau) d\tau
\end{equation}

\noindent
with $x(0)$ and  $v(0)\left(=\dot{x}(0)\right)$ being the initial quantum mechanical
mean values of the co-ordinate and velocity of the particle, respectively. $M_0(t)$
is the inverse form of the Laplace transform of

\begin{equation}
\label{eq25}
\tilde{M}_0(s) = \frac{1}{s^2 + s \tilde{\beta} (s) +\omega_0^2}
\end{equation}

\noindent
with

\begin{equation}
\label{eq26}
\tilde{\beta} (s) = \int_0^\infty \beta(t) e^{-st} dt
\end{equation}

\noindent
is the Laplace
transform of the dissipative kernel $\beta(t)$. The subscript $0$ in $Q_0, M_0$ 
and $G_0$ signifies that the corresponding dynamical  quantities are to be 
calculated around $ x = x_0$.
The time derivative of Eq.(\ref{eq21}) gives

\begin{equation}
\label{eq27}
v(t) = \langle v(t)\rangle_s + \int_0^t m_0(t-\tau) f(\tau) d\tau
\end{equation}

\noindent
where

\begin{equation}
\label{eq28}
\langle v(t)\rangle_s = v(0) m_0(t) -x(0) \omega_o^2 M_0(t) +g_0(t)
\end{equation}

\noindent
with

\begin{equation}
\label{eq29}
m_0(t) = \frac{d}{dt} M_0(t) \; \; {\rm and} \; \; \; 
g_0 = \frac{d}{dt} G_0(t)
\end{equation}

\noindent
It is not difficult to check that $M_0(t)$ and $m_0(t)$ are the two relaxation 
functions; $m_0$ measures how the system with a quantum mechanical mean 
velocity forgets its initial value
while $M_0(t)$ concerns  the relaxation of quantum mechanical mean 
displacement.

Now using the symmetry properties of the correlation function
$\langle f(t) f(t')\rangle_s \left[=c(t-t') = c(t'-t)\right]$ and the solution for
$x(t)$ and $v(t)$ from (\ref{eq21}) and (\ref{eq27}) we obtain the following expressions for the variances

\begin{mathletters}
\begin{eqnarray}
\sigma_{xx}^2 (t) &= &\langle \left[x(t) -\langle x(t)\rangle_s\right]^2\rangle_s \nonumber\\
                  &=&2\int_0^t M_0(t_1) dt_1 \int_0^{t_1} M_0(t_2) c(t_1-t_2) dt_2
\end{eqnarray}

\begin{eqnarray}
\sigma_{vv}^2 (t) &=& \langle \left[v(t) -\langle v(t)\rangle_s\right]^2\rangle_s \nonumber\\
                  &=&2\int_0^t m_0(t_1) dt_1 \int_0^{t_1} m_0(t_2) c(t_1-t_2) dt_2
\end{eqnarray}

\begin{eqnarray}
\sigma_{xv}^2 (t) &=& \langle \left[x(t) -\langle x(t)\rangle_s\right]
\left[v(t) -\langle v(t)\rangle_s\right]\rangle_s \; \; =\frac{1}{2} \dot{\sigma}_{xx}(t) \nonumber\\
                  &=&\int_0^t M_0(t_1) dt_1 \int_0^{t_1} m_0(t_2) c(t_1-t_2) dt_2
\end{eqnarray}
\end{mathletters}

The expressions for variances are general and valid for arbitrary temperature and
friction and includes quantum effects. To recover classical limit of the variances
$\sigma_{xv}^2(t), \sigma_{vv}^2(t), \sigma_{xv}^2(t)$ one has to use (\ref{eq20})
instead of (\ref{eq18}) in 30(a-c). It must also be emphasized that 30(a-c)
are the expressions for statistical variances of the quantum mechanical
mean values $x(t)$ and $v(t)$ with distribution (\ref{eq12}). 
These are not be confused with standard quantum mechanical
variances which are connected through uncertainty relation.

Having obtained the expressions for statistical averages and variances we are now
in a position to write down the quantum Kramers' equation which is a 
Fokker-Planck description for the evolution of true
probability density 
function $p(x, v, t)$ of the quantum mechanical mean values of co-ordinate
and momentum of the particle. To this  end it is necessary to consider the 
statistical distribution of noise $f(t)$ which we assume here to be Gaussian.
For Gaussian noise processes we define \cite{mazo,bjw,maso}
the joint characteristic function 
$\tilde{p}(\mu, \rho, t)$ in terms of the standard mean values and variances,

\begin{equation}
\label{eq31}
\tilde{p}(\mu, \rho, t) = {\rm exp}\left[i\mu \langle x(t) \rangle_s 
+i\rho \langle v(t) \rangle_s -\frac{1}{2}\{\sigma_{xx}^2 \mu^2 + 2 \sigma_{xv}^2
\rho \mu + \sigma_{vv}^2 \rho^2\}\right] \; \; .
\end{equation}

Using standard procedure \cite{adel,mazo,bjw,maso}
we write down the quantum Kramers' equation obeyed by the
joint probability distribution $p(x, v, t)$ which is the inverse Fourier transform
of the characteristic function;

\begin{eqnarray}
\frac{\partial p(x, v, t)}{\partial t} & = & 
\left \{- v\frac{\partial}{\partial x}  +\tilde{V}'(x)\frac{\partial}{\partial v} 
+(\Omega_0(t) 
-N_0(t))\frac{\partial}{\partial v}\right \}p(x, v, t) \nonumber\\
&& + \left \{g_0(t) \frac{\partial}{\partial x} + \tilde{\gamma_0}(t)\frac{\partial}{\partial v}v 
+\phi_0(t)\frac{\partial^2}{\partial v^2} 
+\psi_0(t) \frac{\partial^2}{\partial v \partial x} \right \} p(x, v, t).
\label{eq32}
\end{eqnarray}

\noindent
where

\begin{mathletters}
\begin{equation}
\label{eq33a}
g_0(t) = \dot{G}_0(t) 
\end{equation}

\begin{equation}
\label{eq33b}
\tilde{\gamma}_0(t) = - \frac{d}{dt}\left[ln Y_0(t)\right] 
\end{equation}

\begin{equation}
\label{eq33c}
Y_0(t) = \frac{m_0(t)}{\omega_0^2}\left[1-\omega_0^2 \int_0^t M_0(\tau) d\tau\right] +M_0^2(t) 
\end{equation}

\begin{equation}
\label{eq33d}
\tilde{\omega}_0^2(t) = \frac{1}{Y_0(t)}\left[-M_0(t) \dot{m}_0(t)+m_0^2(t)\right] 
\end{equation}

\begin{equation}
\label{eq33e}
N_0(t) = \frac{1}{Y_0(t)}\left[-g_0(t) \dot{m}_0 \frac{1}{\omega_0^2}
(1-\omega_0^2 \int_0^t M_0(\tau)d\tau)+ m_0^2 G_0)(t)\right] 
\end{equation}

\begin{equation}
\label{eq33f}
\Omega_0(t) = M_0(t) \frac{d}{dt}\left[G_0(t) m_0(t)\right] 
\end{equation}

\begin{equation}
\label{eq33g}
\phi_0(t) = \tilde{\omega}_0^2(t) \sigma_{xv}^2(t) +\tilde{\gamma}_0(t)
\sigma_{vv}^2(t) +\frac{1}{2}\dot{\sigma}_{vv}^2(t) 
\end{equation}

\begin{equation}
\label{eq33h}
\psi_0(t) = \dot{\sigma}_{xv}^2(t) +\tilde{\gamma}_0(t)
\sigma_{xv}^2(t)+ \tilde{\omega}_0^2(t) \sigma_{xx}^2(t)-{\sigma}_{vv}^2(t)  
\end{equation}
\end{mathletters}

$\tilde{V}(x)$ is the renormalized potential linearised at the bottom of the
well at $x=0$, the frequency being $\tilde{\omega}_0(t)$ as given by 
(\ref{eq33d}).
The above Kramers' equation(\ref{eq32}) is the quantum mechanical version of the classical
non-Markovian Kramers' equation and is valid for arbitrary temperature and
friction. The quantum effects appear in the description in two different ways.
First, because of the explicit $Q$-dependence $g_0(t)$ 
[see Eqs.(\ref{eq23}) and (\ref{eq33a})],
$\Omega_0(t)$ and $N_0(t)$ manifestly include the effect of quantum dispersion
of the system through the nonlinearity of the potential. Second, the quantum diffusion
coefficients $\phi_0(t)$ and $\psi_0(t)$ are due to the quantum mechanical
heat reservoir. In the classical limit $g_0(t), \Omega_0(t), N_0(t)$ 
vanishes while $\phi_0(t)$ and $\psi_0(t)$ reduce to the forms which can be 
obtained by using the classical fluctuation-dissipation relation (\ref{eq20}) in (30) 
and (\ref{eq33g}, \ref{eq33h}). 
In the classical limit $k_bT>>\hbar \omega$ Eq.(\ref{eq32})
therefore reduces exactly to non-Markovian Kramers' equation derived earlier by
Adelman and Mazo in late seventies \cite{adel,mazo}.

It is important to emphasize that Eq.(\ref{eq32}) retains its full validity in the quantum limit when $T\rightarrow 0$.
It is also apparent that $G_0$ in Eq. (\ref{eq23}) (therefore in $g_0(t)$) demonstrates
a direct convolution of the relaxation function $M_0(t)$ with quantum dispersion $Q_0(t)$.
This is a clear signature of the interplay of dissipation with nonlinearity of the
potential within a quantum description.

The decisive advantage of the present approach is noteworthy. We have mapped
the operator generalized Langevin equation (\ref{eq3}) into a generalized equation for 
$c$-numbers (\ref{eq16}) and  a corresponding non-Markovian Kramers' equation(\ref{eq32}). The
present approach thus bypasses the earlier methods of quasi-probability distribution
functions employed widely in quantum optics over the decades in a number of
ways. First unlike the {\it quasi-probability} distribution function, the
probability distribution function $p(x, v, t)$ is valid for non-Markovian processes.
Second, while the  corresponding characteristic functions for probability
distribution functions are operators, we make use of the classical characteristic
functions. Third, as pointed out earlier the quasi-distribution functions often
become negative or singular in the  strong quantum domain and pose serious
problems \cite{loui,lnp}. 
The present approach is free from such shortcomings since $p(x, v, t)$
is a {\it true probability distribution} function rather than a 
{\it quasi-probability} function \cite{epw,qpdf}. 
Fourth, the generalized quantum Kramers' equation 
derived here is valid for arbitrary temperature and friction.

Regarding quantum generalized Kramers' equation we further note that although
bounded the time dependent functions $\tilde{\gamma}_0(t), \phi_0(t)$,
$\psi_0(t)$ may not always provide the long time limits. This is well-known in
classical theories \cite{adel,mazo}. These play an important role in the calculation of
non-Markovian Kramers' rate. Therefore, in general, one has to work out the frequency
$\tilde{\omega}_0(t)$ and friction $\tilde{\gamma}_0(t)$ functions for the 
analytically
tractable models. In Sec V we shall consider one such explicit example.

We now consider the stationary distribution of the particles near $x=0$ 
which can be expressed as a solution of Eq.(\ref{eq32}) for 
$\partial p_{st}^0 / \partial t = 0$

\begin{eqnarray}
\left \{v\frac{\partial}{\partial x} -g_0(\infty)
\frac{\partial}{\partial x} -\tilde{\omega}_0^2(\infty) x
\frac{\partial}{\partial v} 
-(\Omega_0(\infty) -N_0(\infty))\frac{\partial}{\partial v}\right \}p_{st}^0(x, v) \nonumber\\
-\left\{\tilde{\gamma_0}(\infty)\frac{\partial}{\partial v}v +\phi_0(\infty)
\frac{\partial^2}{\partial v^2} 
+\psi_0(\infty) \frac{\partial^2}{\partial v \partial x} \right \}p_{st}^0(x, v) = 0 .
\label{eq34}
\end{eqnarray}

\noindent
where the drift and diffusion coefficients of Eq.(\ref{eq34}) assume their 
asymptotic values.

It may be checked immediately that the stationary solution of 
Eq.(\ref{eq34}) is given by

\begin{equation}
\label{eq35}
p_{st}^0(x, v) = \frac{1}{Z}{\rm exp}\left[-\frac{(v-g_0)^2}{2D_0}\right] \times
{\rm exp}\left[-\frac{\tilde{V}(x)+ x(\Omega_0-N_0+\tilde{\gamma}_0g_0)}{D_0+\psi_0}\right]
\end{equation}

\noindent
where $D_0 = \phi_0(\infty) / \tilde{\gamma}_0(\infty); \; \psi_0, \phi_0,
\Omega_0, N_0$ and $g_0$ are the values of the corresponding quantities in the long
time limit. $Z$ is the normalization constant. Here $\tilde{V}(x)$ is the  
renormalized linear potential with a renormalization in its frequency.

Eq. (\ref{eq35}) is the quantum steady state distribution. It may be checked easily that
in the classical Markovian limit the ratio $D_0$ goes over to $k_bT$ while $\psi_0$
vanishes along with $g_0, \Omega_0, N_0$ reducing (\ref{eq35}) to the form of 
Maxwell-Boltzmann distribution function. In what follows in the next section
we shall make use of the quantum
distribution (\ref{eq35}) as a boundary condition for calculation of Kramers' rate.

\section{The quantum Kramers' rate}

We now turn to the problem of barrier crossing dynamics. In Kramers' approach
the particle coordinate $x$ (which in our case it is the quantum mechanical mean 
position) corresponds to the reaction co-ordinate and its values at the 
minima of $V(x)$ denotes the reactant and the product states separated by a
finite barrier, the top being a metastable state representing the transition
state.

Linearizing the motion around the barrier top at $x=x_b$ the Langevin equation can
be written down as

\begin{equation}
\ddot{x} -\omega_b^2 (x - x_b)
+ \int_0^t \beta (t-t')  \dot{x}(t') dt'= f(t) + Q_b(t)
\end{equation}

\noindent
where the barrier frequency $\omega_b^2$ is defined by $V(x) = V(x_b)
-(1/2) \omega_b^2 (x-x_b)^2$. Also the quantum dispersion $Q_b$ has  to be
calculated at the barrier top. Correspondingly the motion of the quantum particle
is governed by the Fokker-Planck equation

\begin{eqnarray}
\frac{\partial p(x, v, t)}{\partial t} & = &
-\left\{ v\frac{\partial}{\partial x} 
-g_b(t) \frac{\partial}{\partial x} 
+\tilde{\omega}_b^2 (x-x_b) \frac{\partial}{\partial v} 
-[\Omega_b(t) -N_b(t)]\frac{\partial}{\partial v}\right\}p(x, v, t) \nonumber\\
&& +\left \{\tilde{\gamma_b}(t)\frac{\partial}{\partial v}v +\phi_b(t)\frac{\partial^2}{\partial v^2}  
+\psi_b(t) \frac{\partial^2}{\partial v \partial x}\right\} p(x, v, t).
\end{eqnarray}

\noindent
where the suffix `$b$' indicates that all the  coefficients are to be calculated at the barrier
top using the general definition of the last section where

\begin{equation}
\tilde{M}_b(s) = \frac{1}{s^2 + s \tilde{\beta} (s) -\omega_b^2}
\end{equation}

\noindent
which is the Laplace transform of $M_b(t)$ and
\begin{mathletters}
\begin{equation}
\chi_x^b(t) = 1+ \omega_b^2 \int_0^t M_b(t') dt'
\end{equation}

Furthermore we have 

\begin{equation}
m_b = \dot{M}_b$ and $g_b(t) = \dot{G}_b(t)
\end{equation}
\end{mathletters}

\begin{mathletters}

\begin{equation}
\tilde{\gamma}_b(t) = - \frac{d}{dt}\left[\ln Y_b(t)\right] 
\end{equation}

\begin{equation}
Y_b(t) = \frac{m_b(t)}{\omega_b^2}\left[1+\omega_b^2 \int_0^t M_b(\tau) d\tau\right] +M_b^2(t) 
\end{equation}

\begin{equation}
\tilde{\omega}_b^2(t) = \frac{1}{Y_b(t)}\left[-M_b(t) \dot{m}_b(t)+m_b^2(t)\right] 
\end{equation}

\begin{equation}
N_b(t) = \frac{1}{Y_b(t)}\left[g_b(t) \dot{m}_b \frac{1}{\omega_b^2}
(1+\omega_b^2 \int_0^t M_b(\tau)d\tau)+ m_b^2 G_b)(t)\right] 
\end{equation}

\begin{equation}
\Omega_b(t) = M_b(t) \frac{d}{dt}\left[G_b(t) m_b(t)\right] 
\end{equation}

\begin{equation}
\phi_b(t) = \tilde{\omega}_b^2(t) \sigma_{xv}^2(t) +\tilde{\gamma}_0(t)
\sigma_{vv}^2(t) +\frac{1}{2}\dot{\sigma}_{vv}^2(t) 
\end{equation}

\begin{equation}
\psi_b(t) = \dot{\sigma}_{xv}^2(t) +\tilde{\gamma}_b(t)
\sigma_{xv}^2(t)+ \tilde{\omega}_b^2(t) \sigma_{xx}^2(t)-{\sigma}_{vv}^2(t)  
\end{equation}
\end{mathletters}

In the spirit of classical Kramers' ansatz \cite{kram} we now demand a solution of Eq. (37)
at the stationary limit of the type.

\begin{equation}
p_{st}(x, v) = p_0(x, v) \zeta(x, v)
\end{equation}

\noindent
with

\begin{equation}
p_0(x, v) = {\rm exp}\left[-\frac{(v-g_b)^2}{2D_b}
-\frac{\tilde{V}(x)+ x(\Omega_b-N_b+\tilde{\gamma}_bg_b)}{D_b+\psi_b}\right]
\end{equation}

\noindent
where $D_b=\phi_b (\infty)/ \tilde{\gamma}_b (\infty); \;
\psi_b, g_b, N_b, \tilde{\gamma}_b$
are the long time limits of the corresponding time dependent quantities specific for
the barrier region. The exponential factor in (41) is not the Boltzmann distribution
but pertains to the dynamics at the barrier top at $x=x_b$. Due to the presence of
$g_b, \Omega_b, N_b$ one may easily comprehend the signature of quantum nature of
the system while $\phi_b, \psi_b$ carries the effect of quantum noise due to heat bath.
The distribution $p_0$ remains finite even at absolute zero.

Now inserting (41), in (37) in the steady state we obtain

\begin{center}
\begin{eqnarray}
-(1+\frac{\psi_b}{D_b})(v-g_b)\frac{\partial \zeta}{\partial x} 
&-&\left[ \frac{D_b}
{\psi_b+D_b} \tilde{\omega}_b^2 
\left(x - x_b -\frac{(\Omega_b-N_b)-\tilde{\gamma}_bg_b)}
{\tilde{\omega}_b^2}\right)+\tilde{\gamma}_b(v-g_b)\right]
\frac{\partial \zeta}{\partial v} \nonumber\\
+\phi_b \frac{\partial^2 \zeta}{\partial v^2} +\psi_b \frac{\partial^2 \zeta}{\partial x \partial v}& =&0
\end{eqnarray}
\end{center}

We then set

\begin{equation}
u = a (x+\alpha_b)+v-g_b
\end{equation}

\noindent
where

\begin{equation}
\alpha_b =-\left[
\frac{\Omega_b-N_b+\tilde{\gamma}_b g_b + x_b \tilde{\omega}_b^2
}{
\tilde{\omega}_b^2}\right] 
\end{equation}

\noindent
and with the help of transformation (44) Eq.(43) is reduced to the following form:

\begin{equation}
(\phi_b+a \psi_b)\frac{\partial^2 \zeta}{\partial u^2} -\left[ \frac{D_b}
{\psi_b+D_b} \tilde{\omega}_b^2(x+\alpha_b)+\{\tilde{\gamma}_b
+a(1+\frac{\psi_b}{D_b})\}(v-g_b)\right]\frac{\partial \zeta}{\partial u} =0
\end{equation}

Now let

\begin{equation}
\frac{D_b}{\psi_b+D_b} \tilde{\omega}_b^2(x+\alpha_b)+\{\tilde{\gamma}_b
+a(1+\frac{\psi_b}{D_b})\}(v-g_b) = \lambda u
\end{equation}

\noindent
where $\lambda$ is a constant to be determined.

From Eq.(44) and (47) we obtain

\begin{equation}
a_{\pm} =-\frac{B}{2A}\pm\sqrt{\frac{B^2}{4A}+\frac{C}{A}}
\end{equation}

\noindent
where
\begin{equation}
A=1+\frac{\psi_b}{D_b}, \; \; B=\tilde{\gamma}_b\;\;\; {\rm and} \; \; \; \;
C=\frac{D_b}{\psi_b+D_b} \tilde{\omega}_b^2
\end{equation}

By virtue of the relation (47) Eq.(46) becomes

\begin{equation}
\frac{d^2 \zeta}{d u^2} +\Lambda u \frac{d \zeta}{d u} =0  \; \; ,
\end{equation}

\noindent
where

\begin{equation}
\Lambda = \left[\frac{\lambda}{\phi_b+a\psi_b}\right]
\end{equation}

The general solution of the homogeneous differential equation (50) is

\begin{equation}
\zeta(u) = F_2\int_0^u {\rm exp}(-\frac{\Lambda u^2}{2}) du +F_1 \; \; \;,
\end{equation}

\noindent
where $F_1$ and $F_2$ are the two constants of integration.

The integral in Eq.(52) converges for $|u|\rightarrow \infty$ if only $\Lambda$
is positive. The positivity of $\Lambda$ depends on the sign of $a$; so by virtue
of Eqns.(44) and (47) we find that the negative root of $a$, i. e., $a_{-}$ 
guarantees
the positivity of $\Lambda$ since $-\lambda a =C$. To determine the value of 
$F_1$ and $F_2$ we impose the first boundary condition on $\zeta$

\begin{equation}
\zeta(x,v)\rightarrow 0 \; \; {\rm as} \; \; x\rightarrow \infty 
\; \; {\rm for\;all} \; \; \; v
\end{equation}

This condition yields

\begin{equation}
F_1 = F_2 \left(\frac{\pi}{2\Lambda}\right)^{1/2}
\end{equation}

By insertion of (54) in (52) we obtain

\begin{equation}
\zeta(u) = F_2 \left[\left(\frac{\pi}{2\Lambda}\right)^{1/2} +\int_0^u e^{-\frac{\Lambda u^2}{2}} du\right]
\end{equation}

Since we are to calculate the current at the barrier top,
we expand the renormalized potential $\tilde{V}(x)$ around $x=x_b$

\begin{equation}
\tilde{V}(x) = \tilde{V}(x_b) -\frac{1}{2}\tilde{\omega}_b^2 (x-x_b)^2
\end{equation}

Thus with the help of (55) and (56) Eq.(41) becomes;

\begin{eqnarray}
p_{st}(x=x_b, v) & =& F_2 {\rm exp} \left(-\frac{\tilde{V}(x_b)+
x_b(\Omega_b-N_b+g_b \tilde{\gamma}_b)}{D_b+\psi_b} \right ) \times \nonumber\\
&& \left[ \left(\frac{\pi}{2\Lambda}\right)^{1/2} {\rm exp}(-\frac{(v-g_b)^2}{2D_b})
+ F(x=x_b, v) {\rm exp}(-\frac{(v-g_b)^2}{2D_b})\right]
\end{eqnarray}

\noindent
with
\begin{equation}
F(x, v) = \int_0^u e^{-\frac{\Lambda u^2}{2}} du
\end{equation}

We now define the steady state  current $j$ across the barrier as

\begin{equation}
j = \int_{-\infty}^{+\infty} v p_{st}(x=x_b, v) dv
\end{equation}

An explicit evaluation of the integral using (57) yields the expression for current
$j$ at the barrier by

\begin{eqnarray}
j & = & F_2 \left[D_b \sqrt{\frac{2\pi D_b}{(1+D_b)}} 
{\rm exp}\{-\frac{\Lambda a^2 (\alpha_b+x_b)^2}
{2(1+\Lambda D_b)}\}+g_b\left\{\left(\frac{\pi}{2\Lambda}\right)^{\frac{1}{2}} \sqrt{2D_b \pi}
+I\right\}\right]\times \nonumber\\
&& {\rm exp}\left\{-\frac{\tilde{V}(x_b)+x_b(\Omega_b-N_b+g_b\tilde{\gamma}_b)}
{D_b+\psi_b}\right\}  
\end{eqnarray}

\noindent
where

\begin{equation}
I = \int_{-\infty}^{+\infty} F(x=x_b, v)\times {\rm exp}[-\frac{(v-g_b)^2}{2D_b}] dv
\end{equation}

Having obtained the stationary current at the barrier top we now determine the 
constant $F_2$ in Eq.(60) in terms of the population of the left well around 
$x=x_0$. This may be done by matching the two appropriate reduced probability
distributions at the bottom of the left well.

To this end we return to Eq.(41) which describes the steady state distribution 
at the barrier top. With the help of (55) we write

\begin{eqnarray}
p_{st}(x, v) & = & F_2 \; {\rm exp}\left[-\frac{\tilde{V}(x)+x(\Omega_b-N_b
+g_b\tilde{\gamma}_b)}{D_b+\psi_b}\right] \times
{\rm exp} \left[-\frac{(v-g_b)^2}{2D_b}\right]\times \nonumber\\
&& \left[\left(\frac{\pi}{2\Lambda}\right)^{1/2}+\int_0^u 
{\rm exp}(-\frac{\Lambda u^2}{2})du \right] 
\end{eqnarray}

We first note that as $x\rightarrow \infty, u\rightarrow \infty$ the 
preexponential factor in $p_{st}(x, v)$ reduces to the form

\begin{equation}
F_2[\cdots] = F_2 \left(\frac{2\pi}{\Lambda}\right)^{1/2}
\end{equation}

We now define a reduced distribution function in $x$ as 

\begin{equation}
\tilde{p}_{st}(x) = \int_{-\infty}^{+\infty} p_{st}(x, v) dv
\end{equation}

Hence from (63) and (64) we obtain

\begin{equation}
\tilde{p}_{st}(x) = 2 \pi F_2 \left(\frac{D_b}{\Lambda}\right)^{1/2}   
\; \times \; {\rm exp} \left [-\frac{\tilde{V}(x)+x(\Omega_b-N_b
+g_b\tilde{\gamma}_b)}{D_b+\psi_b}\right]
\end{equation}

Similarly we derive the reduced distribution function in the left well around
$x= x_0$ using(\ref{eq35}) as
($x_0$ may be put zero without any loss of generality)

\begin{equation}
\tilde{p}_{st}^0(x_0) = \frac{1}{Z} \sqrt{2 \pi D_0}  
\; \times \; {\rm exp} \left[-\frac{\tilde{V}(x_0)+x_0(\Omega_0-N_0
+g_0\tilde{\gamma}_0)}{D_0+\psi_0}\right]
\end{equation}

\noindent
where we have employed the expansion of $\tilde{V}(x)$ as 
$\tilde{V}(x)=\tilde{V}(x_0)+(1/2) \tilde{\omega}_0^2(x-x_0)^2$ and $Z$ is the
normalization constant.

We impose the second boundary condition that at $x=x_0$ the reduced distribution
(65) must coincide with (66) at the bottom of the left well i. e.,

\begin{equation}
\tilde{p}_{st}(x_0) = \tilde{p}_{st}^0(x_0) 
\end{equation}

The above condition is used to determine $F_2$ in terms of normalization
constant $Z$ of (\ref{eq35}).

\begin{eqnarray}
F_2 & = &\frac{1}{Z} \left(\frac{\Lambda}{2\pi}\right)^{1/2}
\left(\frac{D_0}{D_b}\right)^{1/2} 
\; \times \; {\rm exp} \left[-\frac{\tilde{V}(x_0)+x_0(\Omega_0-N_0
+g_0\tilde{\gamma}_0)}{D_0+\psi_0}\right]  \times \nonumber\\
&&  \exp \left[ \frac{\tilde{V}(x_b)-\frac{1}{2}\tilde{\omega}_b^2 
(x_0-x_b)^2+x_0(\Omega_b-N_b
+g_b\tilde{\gamma}_b)}{D_b+\psi_b} \right]
\end{eqnarray}

Furthermore by explicit evaluation of the normalization constant  using the 
integral

\begin{equation}
\int_{-\infty}^{+\infty}\int_{-\infty}^{+\infty} p_{st}^0(x, v) dx dv =1
\end{equation}

\noindent
where $p_{st}^0(x, v)$ is given by (\ref{eq35}). We obtain

\begin{eqnarray}
Z & = & \frac{2\pi}{\tilde{\omega}_0} D_0^{1/2} \left(D_0+\psi_0\right)^{1/2} \; \times
{\rm exp} \left[\frac{[\Omega_0-N_0+g_0\tilde{\gamma}_0]^2}
{2(D_0+\psi_0)\tilde{\omega}_0^2}\right] \times \nonumber\\
&& {\rm exp} \left[-\frac{\tilde{V}(x_0)+x_0(\Omega_0-N_0
+g_0\tilde{\gamma}_0)}{D_0+\psi_0}\right]
\end{eqnarray}

\noindent
Making use of (70) in (68) we obtain from (60) the final expression for 
quantum Kramers' rate, based on flux-over-population method
\cite{hangg,farkas}, as
\begin{eqnarray}
k & = &  \frac{\tilde{\omega}_0}{2 \pi}
\left ( \frac{\Lambda}{2 \pi} \right )^{1/2} 
\frac{1}{D_b^{1/2} \left( D_0 + \psi_0 \right)^{1/2} } \; 
{\rm exp} \left ( - \; 
\frac{ [ \Omega_0 - N_0 + g_0 \tilde{\gamma}_0 ]^2 }{
2 ( D_0 + \psi_0 ) \tilde{\omega}_0^2 } \right ) \nonumber \\
& & \times \left \{ D_b 
\sqrt{ 
\frac{2 \pi D_b}{ 1 + \Lambda D_b } 
} \; {\rm exp} \left ( - \;
\frac{ \Lambda a^2 ( \Omega_b - N_b + g_b \tilde{\gamma}_b )^2 }{
2 ( 1 + \Lambda D_b ) \tilde{\omega}_b^4 } \right ) + g_b
\left [  \left ( \frac{\pi}{2 \Lambda} \right )^{1/2}
\sqrt{2 \pi D_b} + I \right ] \right \} \nonumber \\
& & \times {\rm exp} \left ( - \;
\frac{
E + ( \Omega_b - N_b + g_b \tilde{\gamma}_b ) 
( \sqrt{2E} / \tilde{\omega}_b ) }{ D_b + \psi_b} \right )
\end{eqnarray}

\noindent
where we have used the relation (56) to obtain 
$\tilde{V}(x_0) =\tilde{V}(x_b)-(1/2) \tilde{\omega}_b^2(x_0-x_b)^2$
and definition of activation energy as $E= \tilde{V}(x_b) -\tilde{V}(x_0)$.

A close look into the definitions 39(a), 40(d), 40(e)  of $\Omega, N$ and 
$g$ in the exponential factors in Eq.(71) reveals that each of them is 
proportional to the
anharmonic correction term of the potential, $V'''(x)$ in the leading order so
that, $[\Omega_0-N_0+g_0\tilde{\gamma}_0]^2 / \tilde{\omega}_0^2 \sim
(V'''(x))^2 / (V''(x))$, 
$[\Omega_b-N_b+g_b\tilde{\gamma}_b]^2 / \tilde{\omega}_b^4 \sim
(V'''(x))^2 / (V''(x))^2 $, 
whereas 
$[\Omega_b-N_b+g_b\tilde{\gamma}_b] / \tilde{\omega}_b \sim
(V'''(x)) / \sqrt{(V''(x))} $. The last ratio being the dominant 
contribution the expression (71) then can be simplified as ($\Omega$ vanishes in the  
long time limit)
\begin{eqnarray}
k & = & \frac{ \tilde{\omega}_0}{2 \pi} 
\left ( \frac{\Lambda}{2 \pi} \right )^{1/2}
\frac{1}{D_b^{1/2} ( D_0 + \psi_0 )^{1/2} }
\left \{ D_b 
\sqrt{ 
\frac{2 \pi D_b}{ 1 + \Lambda D_b } 
} + g_b \left [  \left ( \frac{\pi}{2 \Lambda} \right )^{1/2}
\sqrt{2 \pi D_b} + I \right ] \right \} \nonumber \\
& & \times {\rm exp} \left (
\frac{ ( N_b - g_b \tilde{\gamma}_b ) \sqrt{2 E}
}{ \omega_b ( D_b + \psi_b ) }
\right ) \; {\rm exp} \left ( - \; \frac{E}{D_b + \psi_b } \right ) \; \; .
\end{eqnarray}

The above expression is the quantum Kramers' rate which is a direct 
generalization of classical non-Markovian rate valid for intermediate to strong
damping regime and for arbitrary decaying correlation function and temperature. 
The derived rate thus includes the effect of tunneling in a natural way to
modify the classical rate.

To recover the classical non-Markovian expression from (71) one has to take into
consideration (i) the system concerned quantum correction due to non-linearity 
i.e., $N_b$ and $g_b$ must vanish;
(ii) the heat bath noise related quantities like $D$ and $\psi$ are to be calculated
from the expressions of the variances (30) using $c(t-t')$ in the classical 
limit,
i. e., (\ref{eq20}). Under these two conditions (71) is reduced to the classical
expression

\begin{equation}
k = \frac{ \tilde{\omega}_0 }{ 2 \pi } 
\left ( \frac{\Lambda^c}{1 + \Lambda^c D_b^c} \right )^{1/2} 
\frac{D_b^c}{(D_0^c+\psi_0^c)^{1/2} } \; 
{\rm exp} \left ( - \; \frac{ E }{ D_b^c + \psi_b^c} \right ) \; \; \;,
\end{equation}

\noindent
where the superscript $`c'$ signifies the classical limit of the quantum 
mechanical quantities like $D_0(D_b)$ and $\psi_0(\psi_b)$. 
The above expression is identical in form to one derived for classical
non-Markovian dynamics \cite{ray2,ray3,hang}.

\section{A Specific Example: exponentially correlated memory kernel}
 
The structure of $\beta(t)$ given in (\ref{eq19}) suggests that it is quite general
and a further calculation requires a prior knowledge of the density of modes 
$\rho(\omega)$ of the heat bath oscillators. As a specific case we consider in the
continuum limit

\begin{equation}
k(\omega)\rho(\omega)=\frac{2}{\pi}\frac{\Gamma}{(1+ \omega^2 \tau_c^2)}
\end{equation}

\noindent
so that $\beta(t)$ takes the well known form of an exponentially 
correlated memory kernel \cite{maso}

\begin{equation}
\beta(t)=\frac{\Gamma}{\tau_c} {\rm exp}\left[-\frac{|t|}{\tau_c}\right]
\end{equation}

\noindent
where $\Gamma$ is the damping constant and $\tau_c$ refers to the correlation 
time of the noise. Once we get an explicit expression for $\beta(t)$ 
and its Laplace transform 
$\tilde{\beta} (s) = \Gamma / (1+ s\tau_c) $, 
it is possible
to make use of (\ref{eq25}) to calculate $\tilde{M}(s)$ and the relaxation function
M(t) which for the present case 
are given by \cite{maso}
\begin{equation}
\tilde{M_0}(s)=\frac{s+a_0}{s^3+a_0s^2+b_0+c_0}
\end{equation}

\noindent
with
\begin{eqnarray*}
a_0 = \frac{1}{\tau_c} \; \; , \; \; 
b_0 = \omega_o^2+\frac{\Gamma}{\tau_c} \; \; , \; \;
c_0 = \frac{\omega_0^2}{\tau_c} 
\end{eqnarray*}

\noindent
and
\begin{equation}
M_0(t)=c_1^0e^{-p_0t}+c_2^0e^{-q_0t} \sin(\epsilon t+\alpha_0)
\end{equation}

\noindent
respectively, where
\begin{eqnarray}
p_0  =  -A_0-B_0+\frac{a_0}{3} \; \; , \; \;
q_0 = \frac{1}{2}(A_0+B_0)+\frac{a_0}{3} \; \; , \; \;
\epsilon = \frac{\sqrt{3}}{2}(A_0-B_0) \; \; ,
\nonumber \\
c_1^0  =  \frac{1}{2q^0-p^0-d_0} \; \; , \; \;
d_0  =  \frac{a_0(2q^0-p^0)-{q^0}^2-\epsilon^2}{a_0-p^0} \; \; ,
\nonumber\\
A_0  =  \left ( \frac{-a_0^3}{27}+\frac{a_0b_0}{6}-\frac{c_0}{2}
+\sqrt{R_0} \right )^{1/3} \; \; , \; \;
B_0  =  \left ( -\frac{a_0^3}{27}+\frac{a_0b_0}{6}-\frac{c_0}{2}
-\sqrt{R_0} \right )^{1/3} \; \; , 
\nonumber\\
c_2^0   =  -\frac{c_1^0}{\epsilon}[(d_0-q^0)^2+
\epsilon]^{1/2} \; \; {\rm and} \; \;
\alpha_0  =  \tan^{-1} \left ( \frac{\epsilon}{d_0-q^0} \right )
\end{eqnarray}

\noindent
and 
\begin{eqnarray*}
R_0=-\frac{a_0^2b_0^2}{108}+\frac{b_0^3}{27}+\frac{a_0^3c
_0}{27}-\frac{a_0b_0c_0}{6}+\frac{c_0^2}{4} \; > \; 0 
\end{eqnarray*}

\noindent
for the present problem.

Now making use of the expression (77) for M(t) and expression (\ref{eq18}) for 
correlation function $c(t)$ in Eq.(30(a-c)) we calculate explicitly after a
tedious long but straight forward algebra the time dependent expressions 
for the variances of the quantum mechanical mean values of position and momentum of the particle.
These expressions are given by

\begin{eqnarray*}
\sigma_{xx}^2(t)=\frac{2\hbar \Gamma}{\pi} \int_0^\infty\frac{\omega}{1+\omega
^2\tau_c^2}(\coth{\frac{\hbar\omega}{2k_bT}}){\cal F}_x(\omega,t)d\omega
\nonumber\\\sigma_{vv}^2(t)=\frac{2\hbar \Gamma}{\pi}\int_0^\infty\frac{\omega}
{1+\omega^2\tau_c^2}(\coth\frac{\hbar\omega}{2k_bT}){\cal F}_v(\omega,t)d\omega
\end{eqnarray*}

\noindent 
and

\begin{eqnarray*} 
\sigma_{xv}^2(t)=\frac{1}{2}\dot{\sigma}_{xx}^2(t)
\end{eqnarray*}

In Appendix-B we provide the explicit structures of ${\cal F}_x(\omega, t)$ and 
${\cal F}_v(\omega,t)$. Since in the long time limit $\sigma_{xv}^2, \dot
{\sigma}_{xv}^2, \dot{\sigma}_{vv}^2$ vanish, the calculation of the quantities
$D_0,D_b, \psi_0, \psi_b$ essentially rest on the asymptotic values of $\sigma 
_{xx}^2(t)$ and $\sigma_{vv}^2(t)$ evaluated at the barrier top or the 
bottom of the well. It must be emphasized that the quantities are 
non-vanishing at $T=0$ due to quantum fluctuation of the heat bath.

We now turn to the calculation of the relevant asymptotic coefficients
$g_b$  and  $N_b$ in the expression for the rate ($\Omega_0$ and $\Omega_b$ vanish in the 
long time limit because of the relaxation function). Both of them are related
to the convolution integral through the relations (40d) and $g_b=\dot{G_b}$.
Since $G_b$ is defined as $\int_0^tM_b(t-\tau)Q_b(\tau)d\tau$ where $M_b$
and $Q_b$ correspond to the barrier top  (and $M_0$ and $Q_0$ to the bottom 
of the potential well)  we make use of the expressions for $M_b$ and $M_0$ 
as given by (77) along with those for $Q_b$ and $Q_0$ as shown in Appendix-A
to obtain $G_0(t)$ and $G_b(t)$ and their asymptotic values. Considering only
the short time linearity of $G_b(t)$ (since the quantum effect in 
$Q_b(t)$ has been taken into account for the lowest order in the Appendix-A) 
calculation of
$g_b$ is quite straight forward. Furthermore the expression for $N_b$ as given by 
(40d) can be simplified in the asymptotic limit to obtain 

\begin{eqnarray*} 
N_b(t)=\frac{\dot{M}_b^2G_b(t)}{M_b^2} 
\end{eqnarray*}

Both $g_b$ and $N_b$ involve the constants ${\langle\delta\hat{X}^2\rangle} 
_{t=0}$ and ${\langle\delta\hat{X}\delta\hat{P}+\delta\hat{P}\delta\hat{X}
\rangle}_{t=0}$ which are assumed to be $\hbar / (2\omega_b)$ (minimum
uncertainty state) and zero, respectively for the present calculation.

To analyze the associated non-Markovian nature of the dynamics at various
temperatures it is necessary to go over to numerical simulation of
stationary values of $D_0, D_b$ and $\psi_0, \psi_b$ and $\Lambda$. These
in turn, are primarily dependent in $\sigma_{xx}^2(t)$ and $\sigma_{vv}^2(t)$,
the other variances being vanishing in the long time limit. For the present
purpose we assume the simplest form of the cubic potential of the
type $V(x) = - (1/3) \overline{A} x^3 + \overline{B} x^2$ where
the parameter set used is $\overline{A} = 0.5$; 
$\overline{B} = \left [ (3/4) \overline{A}^2 E \right ]^{1/3}$; the activation
energy $E = V(x_b) - V(x_0) = 10$. The correlation time of the noise $\tau_c$
is fixed at $0.3$. The temperature and the damping constant $\Gamma$ are 
varied set to set.
The quantities $g_b$ and $N_b$ which incorporate quantum effects through the
anharmonicity of the potential can be easily calculated numerically as 
outlined in the previous paragraph. In Fig.~1 we show the 
Arrhenius plot, i. e,
the variation of $\ln k$ vs $1/T$ for two different values of
damping constant, $\Gamma$. It is apparent that in the high temperature
regime the plot exhibits linearity, which is the standard Arrhenius classic
result. In the low temperature regime, however, one observes a much
slower variation which is a typical quantum behaviour. To single out this low
temperature behaviour, we show in the inset of Fig.~1 a clear $T^2$
dependence of the rate - a feature observed earlier in the recent past \cite{hangg}.
In Fig.~2(a-c) we exhibit the variation of the rate $k$ as a function of damping
constant $\Gamma$ at several temperatures. It is apparent that at a
relatively high temperature the rate varies inversely with the damping constant while
at low temperature the rate drops at a much faster rate. At $T = 0$
the decay is exponential in nature. The quantum rate in this situation
essentially corresponds to zero-temperature tunneling. This result is in
satisfactory agreement with that of Caldeira and Leggett \cite{legg}. 
The present theory
therefore unifies the aspects of quantum tunneling and thermal
noise-induced barrier crossing on the same footing.

\section{conclusions}
In this paper we have proposed a simple approach to non-Markovian
theory of quantum Brownian motion in phase space. Based on an initial
coherent state representation of bath oscillators and a canonical 
equilibrium distribution of
quantum mechanical mean values and their coordinates and momenta, we have shown that
it is possible to realize a stochastic differential equation in c-numbers
in the form of a generalized Langevin equation and the associated 
Fokker-Planck
equation which can be recognized as a generalized quantum Kramers' equation. The Kramers'
equation is then employed to derive the rate of barrier crossing 
which includes both tunneling and thermal induced effects on the same footing.
The main conclusions in this study are the following:

(i) Our ensemble averaging procedure and the QGLE are amenable to theoretical 
analysis in terms of the methods developed earlier for the treatment of
classical non-Markovian theory of Brownian motion.

(ii)  The proposed Kramers' equation is an exact quantum analogue of classical
generalized Kramers' equation derived earlier in late seventies by a number
of workers. Since we have dealt here with {\it true probability
functions} the theory is free from the problem of singularity or negativity
of {\it quasi-classical} distribution functions which is often encountered in 
Wigner equation approaches. The equation is valid for arbitrary temperature
and friction.

(iii) The realization of noise as a classical looking entity which satisfies
quantum fluctuation-dissipation relation allows ourselves to envisage 
quantum Brownian motion as a quantum generalization of 
its classical counterpart.
The method is based on the canonical quantization procedure and 
is independent of the path integral formalisms.

(iv)  The quantum Kramers' rate (Eq.71) is valid for intermediate to strong 
damping regime and for arbitrary temperature and decaying noise correlations.
It reduces to classical non-Markovian rate and purely vacuum fluctuation- 
induced rate  or tunneling in the appropriate limits.

(v) The theory incorporates quantum effects in two different ways. The 
quantum nature of the system is manifested through the non-linear part of the
potential while the heat bath imparts the usual quantum noise.
It must be emphasized that our general analysis takes into consideration 
of quantum effects of all orders .

(vi)  The theory also reveals an interesting interplay of non-linearity
and dissipation in the Fokker-Planck coefficients which include quantum
corrections. The variation of the rate due to tunneling and activation with respect
to temperature and damping has been clearly demonstrated.

The theory presented here is a natural extension of the classical theory
of Brownian motion in the sense that the quantum Kramers' equation is {\it classical-looking in form but quantum mechanical in its content}.
Also we have considered only the spatial diffusion limited regime in the
calculation of the rate.
It is worthwhile to extend the approach to the energy diffusion regime and 
further to implement other methods of treatment of classical Brownian motion.

\acknowledgements
S K Banik is indebted to the Council of Scientific and Industrial Research
(C.S.I.R.), Government of India for financial support.

\begin{appendix}

\section{calculation  of quantum dispersion Q}

The quantum fluctuation $Q$ is defined in (\ref{eq15}) as 

\begin{equation}
Q(t)=V'(\langle\hat{X}\rangle)-\langle V'(\hat{X})\rangle
\end{equation}

So far as the general formulation of the theory upto Sec.IV is concerned $Q$ is taken 
in full. Or in other words the quantum Kramers' equation (\ref{eq32}) or the rate Eq.(71)
incorporates quantum effects due to the system in all orders. However in
actual calculations Q has to be estimated order by order \cite{sund}. To this end  
we consider the lowest order quantum corrections. Returning to the quantum
mechanics of the system in Heisenberg picture it is convenient to write the 
operators $\hat{X}$ and $\hat{P}$ as 

\begin{eqnarray}
\hat{X}(t)=\langle\hat{X}(t)\rangle+\delta\hat{X}\nonumber\\\label{eqA2} 
\hat{P}(t)=\langle\hat{P}(t)\rangle+\delta\hat{P}
\end{eqnarray}

\noindent
$\langle\hat{X}(t)\rangle$ and $\langle\hat{P}(t)\rangle$ are the 
operators signifying quantum mechanical  
averages and $\delta\hat{X}$ and $\delta\hat{P}$ are quantum
corrections. By construction $\langle\delta\hat{X}\rangle$ and $\langle\delta
\hat{P}\rangle$ are zero and $\delta\hat{X}$ and $\delta\hat{P}$ obey the
commutation relation $[\delta\hat{X},\delta\hat{P}]=i\hbar$. Using (A2)
in $\langle V'(\hat{X})\rangle$ and a Taylor expansion around $\langle
\hat{X}\rangle$ it is possible to express $Q(t)$ as (keeping the lowest order
non-vanishing term)

\begin{equation}\label{A3}\\Q(t)=-\frac{1}{2}V'''(\langle
\hat{X}\rangle)\langle\delta\hat{X}^2(t)\rangle
\end{equation}

\noindent
where $\langle X\rangle$ and $\langle\delta X^2\rangle$ follow  a coupled set
of equations as given below ;

\begin{eqnarray*}
\langle\dot{\hat{X}}\rangle=\langle\hat{P}\rangle
\end{eqnarray*}

\begin{eqnarray*}
\langle\dot{\hat{P}}\rangle=-V'(\langle\hat{X}\rangle)
\end{eqnarray*}

\begin{eqnarray*}
\frac{d}{dt}\langle\delta{\hat{X}}^2\rangle=\langle\delta\hat{X}\delta\hat
{P}+\delta\hat{P}\delta\hat{X}\rangle
\end{eqnarray*}

\begin{eqnarray*}
\frac{d}{dt}\langle\delta\hat{X}\delta
\hat{P}+\delta\hat{P}\delta\hat{X}\rangle=2\langle\delta\hat{P}^2\rangle-
2V''(\langle\hat{X}\rangle)\langle\delta\hat{X}^2\rangle\nonumber\\
\end{eqnarray*}

\begin{equation}
\frac{d}{dt}\langle\delta{\hat{P}}^2\rangle=-V''(\langle\hat{X}\rangle)\langle\delta
\hat{X}\delta\hat{P}+\delta{P}\delta\hat{X}\rangle
\end{equation}

The above set of equations can be derived from Heisenberg's equation of motion. 
In the Kramers' problem one is primarily concerned with the local dynamics of the
system around the barrier top at $\langle\hat{X}\rangle(=x) =x_b$ or at the  
bottom at $\langle\hat{X}\rangle (=x)=x_0$. The solution depends critically
on the nature of curvature of the potential, i.e., $V''(\langle X\rangle)$.
Considering the local nature of the dynamics we have expressed 
$V''(x=x_b)=-\omega_b^2$ and $V''(x=x_0)=\omega_0^2$ and consequently the equations for the 
quantum corrections can be solved independently of the first two of Eqs.(A4).
A solution of $\langle\delta\hat{X}^2\rangle$ around $x=x_0$ 
which is an elliptic fixed point is therefore 
given by

\begin{equation}
{\langle\delta\hat{X}^2(t)\rangle}_{x \sim x_0}=
\frac{1}{2}[{\langle\delta\hat{X}^2\rangle}_{t=0}-\frac{{\langle\delta\hat{P}^2
\rangle}_{t=0}}{\omega_0^2}]\cos(2\omega_0t)+\frac{{\langle\delta\hat{X}
\delta\hat{P}+\delta\hat{P}\delta{X}\rangle}_{t=0}}{2\omega_0}\sin(2\omega_0t)
+\frac{2I_0^c}{4\omega_0^2}
\end{equation} 

\noindent
where $I_0^c$ is an integration
constant and is given by $I_0^c=\langle\delta\hat{P}^2\rangle+\omega_0^2
{\langle\delta{X}^2\rangle}_{t=0}$. 
Therefore the quantum dispersion at the bottom of the well is given by,

\begin{equation}\label{eqA6}Q_0(t)=-\frac{1}{2}V'''(x_0){\langle\delta{\hat
{X}}^2(t)\rangle}_{x\sim x_0}
\end{equation}

\noindent
where ${\langle\delta{\hat{X}}^2
(t)\rangle}_{x\sim x_0}$ is governed by Eq.(A5).
Similarly we calculate quantum fluctuation $Q_b(t)$ near the top of the
potential barrier at $x= x_b$ as

\begin{equation}\label{eqA7}Q_b(t)=-\frac{1}{2}
V'''(x_b){\langle\delta{\hat{X}}^2(t)\rangle}_{x\sim x_b}
\end{equation}

\noindent
where ${\langle\delta{\hat{X}}^2(t)\rangle}_{x\sim x_b}$ in a solution of 
the last three equation of (A4) around $x\sim x_b$ and is given by

\begin{equation}\label{A8}{\langle\delta{\hat{X}}^2(t)\rangle}_{x\sim x_b}={\langle
\delta{\hat{X}}^2\rangle}_{t=0}\cosh(2\omega_bt)+\frac{{\langle\delta\hat{X}
\delta\hat{P}+\delta\hat{P}\delta\hat{X}\rangle}_{t=0}}{2\omega_b}\sinh(2
\omega_bt)
\end{equation}

It is interesting to note the hyperbolic nature of the top of the
barrier which is reflected in the exponential divergence of the quantum 
fluctuations. This point has been studied extensively in the recent 
literature in the context of chaos \cite{fox,chau,chau1,chau2}. Having evaluated the quantum dispersions
$Q_0(t)$ and $Q_b(t)$ we are now in a position to calculate several
related quantities like $G(t), N(t), \Omega(t)$ and $g(t)$. A better estimate of the quantum correction 
$Q$ can be obtained from the solutions of the equations of higher order
corrections derived earlier by Sundaram and Milonni \cite{sund}.

\section{Calculation  of ${\cal F}_x(\omega, t)$ and ${\cal F}_v(\omega, t)$}

The expressions ${\cal F}_x(\omega,t)$ and ${\cal F}_v(\omega,t)$ are 
given by 

\begin{equation}
{\cal F}_x(\omega,t)=\int_0^tM_0(t_1)
[c_1^0I_1(t_1)+c_2^0I_2(t_1)]dt_1\end{equation} and \begin{equation}\label 
{eqB2}{\cal F}_v(\omega,t)=\int_0^tN_0(t_1)[p^0c_1^0I_!(t_1)+q_0c_2^0I_2(t_1)
-c_2^0\epsilon I_3(t_1)]dt_1
\end{equation}

\noindent
where 

\begin{equation}
N_0(t)=c_1^0p^0e^{-p^0t}+c_2^0q^0e^{-q^0t}\sin(\epsilon 
t+\alpha)-c_2^0\epsilon e^{-q^0t}\cos(\epsilon t+\alpha)
\end{equation}

Here $I_1$, $I_2$, and $I_3$ are given by the following expressions.

\begin{equation}
I_1\;=\;\;-B_{2j}e^{-pt_1}\;+\;B_{2j}\cos\;\omega_jt_1\;+\;B_{1j}\sin\;\omega_jt_1\\
\end{equation}

\begin{eqnarray}
I_2 &=& \frac{1}{2}[-B_{4j}\cos(\omega_jt_1+\alpha)+ B_{4j}e^{-qt_1}\cos(\omega_j
t_1+\alpha)\cos(\omega_{j-}t_1)+B_{6j}e^{-qt_1}\cos(\omega_jt_1+\alpha) \nonumber\\
&& \times \sin(\omega_{j-}t_1) 
+B_{6j}\sin(\omega_jt_1+\alpha)-B_{6j}e^{-qt_1}\sin(\omega
_jt_1+\alpha)\cos(\omega_{j-}t_1)+
 B_{4j}e^{-qt_1} \; \times \nonumber\\
&& \sin(\omega_jt_1+\alpha)\sin(\omega_{j-}t_1) 
+B_{3j}\cos(\omega_jt_1-\alpha)-B_{3j}e^{-qt_1}\cos(
\omega_jt_1-\alpha)\cos(\omega_{j+}t_1)  \nonumber\\
&& -B_{5j}e^{-qt_1}\cos(\omega_jt_1-\alpha)\sin(\omega_{j+}t_1)
+B_{5j}\sin(\omega_jt_1-\alpha)-
B_{5j}e^{-qt_1} \sin(\omega_jt_1-\alpha) \; \times \nonumber\\
&&\cos(\omega_{j+}t_1) 
+B_{3j}e^{-qt_1}\sin(\omega_j
t_1-\alpha)\sin(\omega_{j+}t_1)]
\end{eqnarray}

\begin{eqnarray}
I_3 & = & \frac{1}{2}[B_{6j}\cos(\omega_jt_1
+\alpha)- B_{6j}e^{-qt_1}\cos(\omega_jt_1+\alpha)\cos(\omega_{j-}t_1)+B_{4j}e^{-qt_1}
\cos(\omega_jt_1+\alpha)\nonumber\\
&& \times \sin\;\omega_{j-}t_1\;+\;B_{4j}
\sin(\omega_jt_1+\alpha)-\;B_{4j}e^{-qt_1}\sin(\omega_jt_1+\alpha)\cos(\omega
_{j-}t_1)-\nonumber\\
&& B_{6j}e^{-qt_1}
\sin(\omega_jt_1+\alpha)\sin(\omega_{j-}t_1)+
B_{5j}\cos(\omega_jt_1-\alpha)\;-\;B_{5j}e^{-qt_1}\cos(\omega_jt_1-\alpha) \nonumber\\
&& \times \cos(\omega_{j+}t_1)
+B_{3j}e^{-qt_1}\cos(\omega_jt_1-\alpha)\sin(\omega_{j+}t_1)
+B_{3j}\sin(\omega_jt_1-\alpha)-B_{3j}e^{-qt_1} \nonumber\\ 
&& \times \sin(\omega_jt_1-\alpha) 
 \cos(\omega_{j+}t_1)- B_{5j}e^{-qt_1}\sin(\omega_jt_1-\alpha)\sin(
\omega_{j+}t_1)]
\end{eqnarray}

where
\begin{eqnarray*}
\omega_{j+}=\;\omega_j\;+\;\alpha\;\;\;\;\;\;\;\;\;\omega_{j-}=\;\omega_j\;-\;
\alpha\\B_{1j}=\;\frac{\omega_j}{p^2+{\omega_j}^2}\;\;\;\;\;\;\;\;\;\;\;\;
B_{2j}=\;\frac{p}{p^2+{\omega_j}^2}\\B_{3j}=\;\frac{\omega_{j+}}{q^2+{\omega_
{j+}}^2}\;\;\;\;\;\;\;\;\;\;\;\;B_{4j}=\;\frac{\omega_{j-}}{q^2+{\omega
_{j-}}^2}\\B_{5j}=\;\frac{q}{q^2+{\omega_{j+}}^2}\;\;\;\;\;\;\;\;\;\;\;\;
B_{6j}=\;\frac{q}{q^2+{\omega_{j-}}^2}
\end{eqnarray*}
\end{appendix}

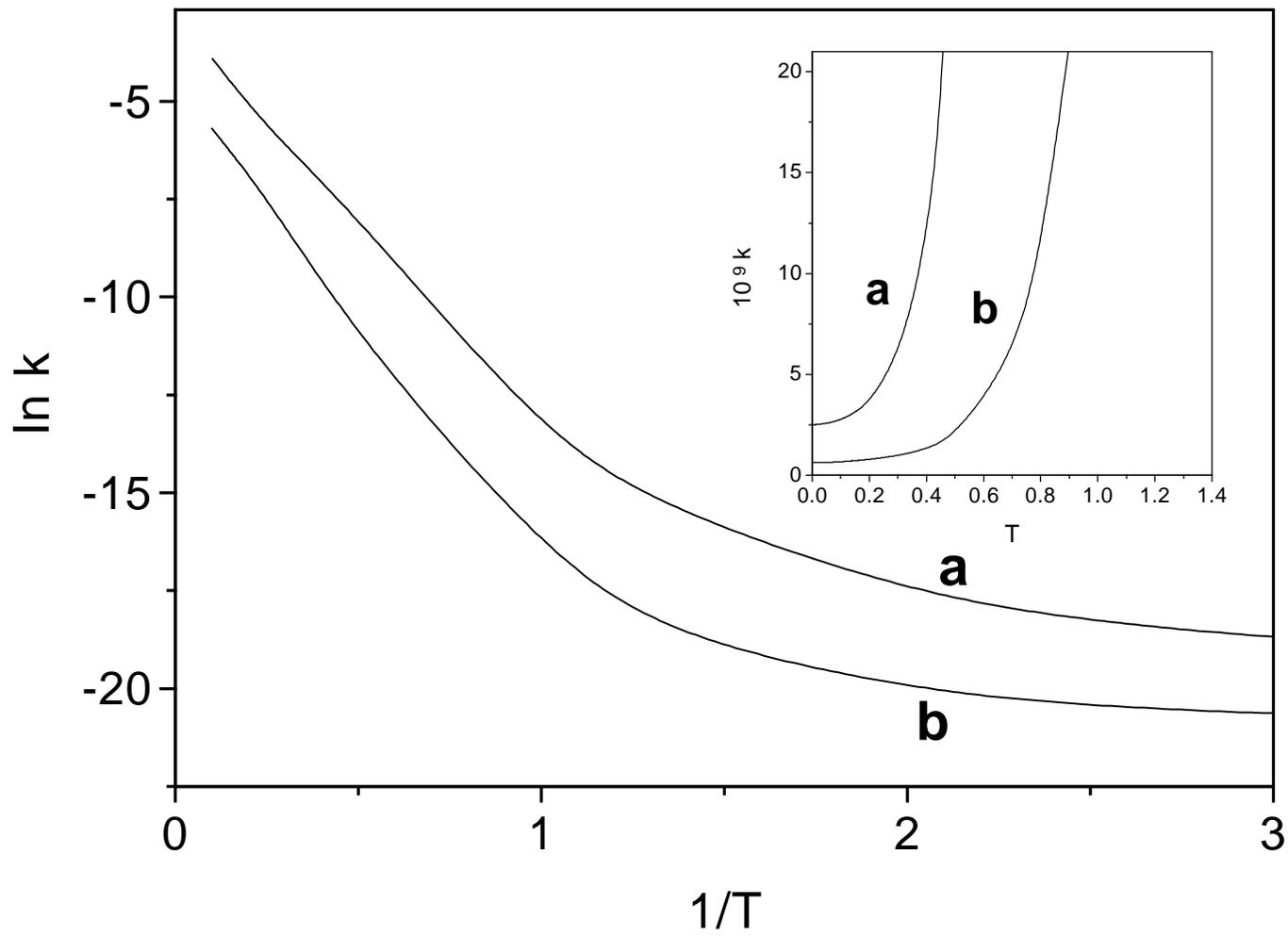
\begin{figure}
\caption{
Plot of $\ln k$ vs $1/T$ using Eq.(72) for different values of
the damping contant (a) $\Gamma =1.3$, and (b) $\Gamma =1.7$ 
[ Inset : Plot of $k$ vs $T$ to illustrate the $T^2$ dependence at very low
temperature. The parameters are same as in the main figure ]
(units arbitrary).
}
\end{figure}

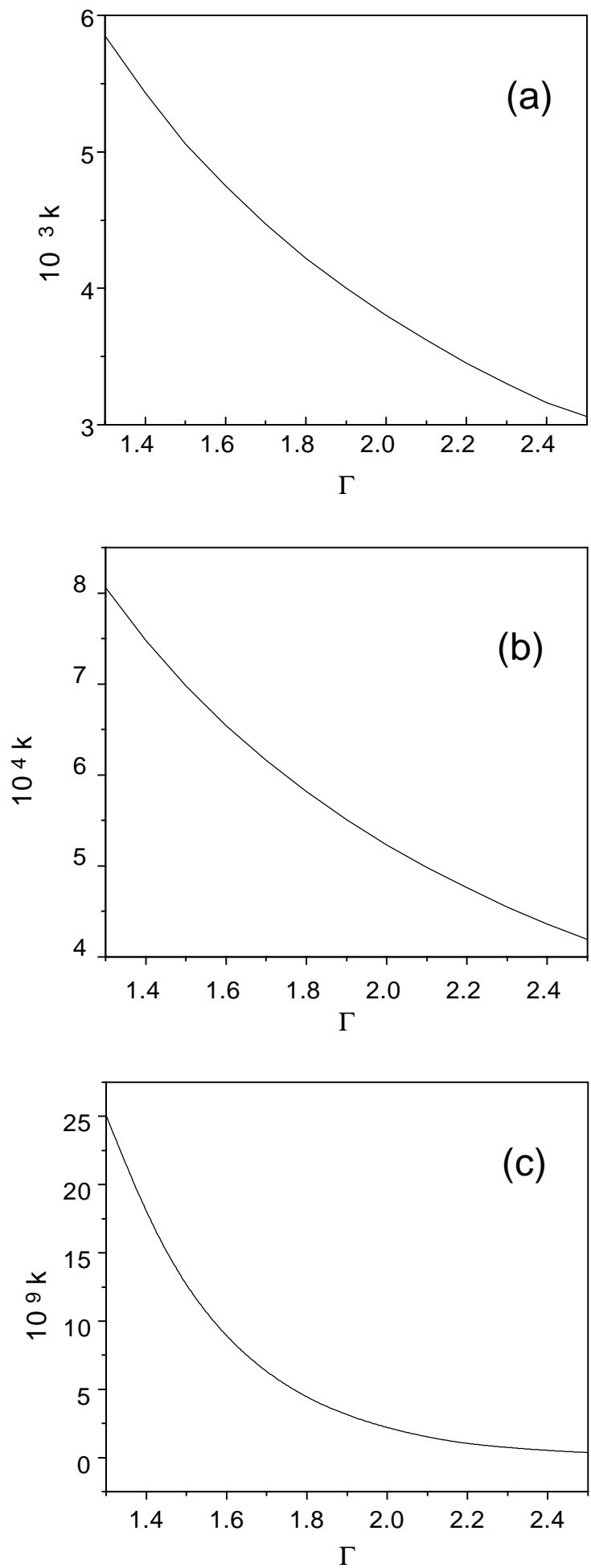
\begin{figure}
\caption{
Plot of quantum Kramers' rate $k$ vs $\Gamma$ using Eq.(72) 
for (a) $T = 5.0$, (b) $T = 3.0$ and (c) $T = 0.0$
(units arbitrary).
}
\end{figure}


\begin{thebibliography}{99}

\bibitem{kram}
H. A. Kramers, Physica (Utrecht) {\bf 7}, 284 (1940).

\bibitem{hangg}
P. H\"anggi, P. Talkner and M. Borkovec, Rev. Mod. Phys. {\bf 62}, 251 (1990).

\bibitem{mel}
V. I. Mel'nikov, Phys. Rep. {\bf 209}, 1 (1991).

\bibitem{tani}
Y. Tanimura and R. Kubo, J. Phys. Soc. Japan {\bf 58}, 1199 (1989).

\bibitem{tani1}
Y. Tanimura and P. G. Wolynes, Phys. Rev. A {\bf 43}, 4131 (1991).

\bibitem{stru}
W. T. Strunz, L. Di\'osi, N. Gisin and T. Yu, Phys. Rev. Lett. {\bf 83}, 
4909 (1999).

\bibitem{joac}
P. Pechukas, J. Ankerhold and H. Grabert, Ann. Phys. (Leipzig) 
{\bf 9}, 794 (2000);
J. Ankerhold, P. Pechukas and H. Grabert, Phys. Rev. Lett. {\bf 87},
086802 (2001).

\bibitem{laio}
F. Laio, A. Porporato, L. Ridolfi  and I. Rodriguez-Iturbe, 
Phys. Rev. E {\bf 63}, 036105 (2001).

\bibitem{mant}
R. N. Mantegna and B. Spagnolo, Phys. Rev. Lett. {\bf 84}, 3025 (2000)

\bibitem{bier}
M. Bier and R. D. Astumian, Phys. Rev. Lett. {\bf 71}, 1649 (1993). 

\bibitem{ray}
J. Ray Chaudhuri, G. Gangopadhyay and D. S. Ray, J. Chem. Phys. {\bf 109},
5565 (1998).

\bibitem{ray1}
J. Ray Chaudhuri, B. C. Bag and D. S. Ray, J. Chem. Phys. {\bf 111},
10852 (1999).

\bibitem{ray2}
S. K. Banik, J. Ray Chaudhuri and D. S. Ray, J. Chem. Phys. {\bf 112},
8330 (2000).

\bibitem{ray3}
J. Ray Chaudhuri, S. K. Banik, B. C. Bag and D. S. Ray, Phys. Rev. E 
{\bf 63}, 061111 (2001).

\bibitem{loui}
W. H. Louisell, {\it Quantum Statistical Properties of Radiation}
(Wiley, New York, 1973).

\bibitem{gang}
G. Gangopadhyay and D. S. Ray, in 
{\it Advances in Multiphoton Processes and Spectroscopy},
Vol.~8, edited by S. H. Lin, A. A. Villayes and F. Fujimura
(World Scientific, Singapore, 1993).

\bibitem{mill}
W. H. Miller, J. Chem. Phys. {\bf 62}, 1899 (1975); P. G. Wolynes, Phys.
Rev. Lett. {\bf 47}, 968 (1981); W. H. Miller, S. D. Schwartz and J. W.
Tromp, J. Chem. Phys. {\bf 79}, 4889 (1983).

\bibitem{cao}
J. Cao and G. A. Voth, J. Chem. Phys. {\bf 105}, 6856 (1996).

\bibitem{liao}
J. Liao and E. Pollak, J. Chem. Phys. {\bf 110}, 80 (1999).

\bibitem{legg}
A. O. Caldeira and A. J. Leggett, Phys. Rev. Lett. {\bf 46}, 211 (1981);
Ann. Phys. (N.Y.) {\bf 149}, 374 (1983);
Physica A {\bf 121}, 587 (1983).

\bibitem{grab}
H. Grabert, P. Schramm and G. L. Ingold, Phys. Rep. {\bf 168}, 115 (1988).

\bibitem{legg1}
A. J. Leggett, S. Chakravarty, A. T. Dorsey, M. P. A. Fisher, A. Garg and
W. Zwerger, Rev. Mod. Phys. {\bf 59}, 1 (1987). 

\bibitem{talk}
P. Talker, E. Pollak and A. M. Berezhkovskii (Eds.), 
Chem. Phys. {\bf 235}, 1 (1998).

\bibitem{adel}
S. A. Adelman, J. Chem. Phys. {\bf 64}, 124 (1976).

\bibitem{mazo}
R. M. Mazo, in {\it Stochastic Processes in Nonequilibrium Systems}, 
edited by L. Garido, P. Segler and P. J. Shepherd, 
Lecture Notes in Physics, Vol.~84 (Springer Verlag, Berlin 1978).

\bibitem{hang}
P. H\"anggi and F. Mojtabai, Phys. Rev. A {\bf 26}, 1168 (1982). 

\bibitem{grot}
R. F. Grote and J. T. Hynes, J. Chem. Phys. {\bf 73}, 2715 (1980).

\bibitem{epw} E. Wigner, Phys. Rev. {\bf 40}, 749 (1932).

\bibitem{qpdf} M. Hillery, R. F. O'Connell, M. O. Scully and E. P. Wigner,
{\bf 106}, 121 (1984).

\bibitem{zwanzig}
R. Zwanzig, J. Stat. Phys. {\bf 9}, 215 (1973).

\bibitem{bjw}
K. Lindenberg and B. J. West, {\it The Nonequilibrium Statistical Mechanics
of Open and Closed Systems} (VCH, New York, 1990).

\bibitem{lnp} H. Risken and K. Vogel in {\it Far from equilibrium phase
transition}, edited by L. Garrido, Lecture Notes in Physics,
Vol.~{319} (Springer-Verlag, Berlin 1988).

\bibitem{farkas} L. Farkas, Z. Phys. Chem. (Leipzig) {\bf 125}, 236 (1927).

\bibitem{maso}
Ke-Gang Wang and J. Masoliver, Physica A {\bf 231}, 615 (1996). 

\bibitem{sund}
B. Sundaram and P. W. Milonni, Phys. Rev. E {\bf 51}, 1971 (1995).

\bibitem{fox}
R. F. Fox and T. C. Elston, Phys. Rev. E {\bf 49}, 3683 (1994).

\bibitem{chau}
S. Chaudhuri, G. Gangopadhyay and D. S. Ray, Phys. Rev. E  {\bf 54},
2359 (1996).

\bibitem{chau1}
B. C. Bag, S. Chaudhuri, J. Ray Chaudhuri and D. S. Ray, Physica D  {\bf 96},
47 (1999).

\bibitem{chau2}
B. C. Bag and D. S. Ray, J. Stat. Phys. {\bf 96}, 271 (1999).


\end{thebibliography}
\end{document}